\documentclass{elsart}
\usepackage{amsmath,amsthm}
\usepackage{makeidx,graphics}
\usepackage{epsfig}
\usepackage{colordvi}

\def\be{\begin{eqnarray*}}
\def\ee{\end{eqnarray*}}

\newcommand{\scr}[1] {\mathcal{#1 }}

\newcommand{\refe}[1]{equation (\ref{#1})}

\newcommand{\beq}{\begin{eqnarray}}
\newcommand{\eeq}{\end{eqnarray}}

\begin{document}
\runauthor{Coppersmith, Kadanoff and Zhang}
\begin{frontmatter}
\title{Reversible Boolean Networks I: Distribution of Cycle Lengths}
\author[Chicago]{S. N. Coppersmith, Leo P. Kadanoff, Zhitong Zhang}
\address[Chicago]{the James Frank Institute, the University of Chicago,
5640 S. Ellis Ave, Chicago, IL 60637}

\begin{abstract}

We consider a class of models describing the dynamics of
$N$ Boolean variables, where the time evolution of each depends
on the values of $K$ of the other variables.
Previous work has considered models with dissipative dynamics.
Here we consider time-reversible
models, which necessarily have the property that every
possible point in the state-space is an element of one
and only one cycle.

As in the dissipative case, when $K$ is large, typical orbit lengths
grow exponentially with $N$, whereas for small enough $K$, typical orbit
lengths grow much more slowly with $N$. The
numerical data are consistent with the existence of a phase transition
at which the average orbit length grows as a power of $N$ at a value of $K$ between
1.4 and 1.7. However, in the reversible models
the interplay between the discrete symmetry and quenched randomness can lead to
enormous fluctuations of orbit lengths and other interesting features
that are unique to the reversible case.

The orbits can be classified by their behavior under
time reversal. The orbits that transform into themselves under time
reversal have properties quite different from those
that do not; in particular, a significant fraction of
latter-type orbits have lengths enormously longer than orbits that are time
reversal-symmetric. For large $K$ and moderate $N$ , the vast majority 
of points in the state-space are on one of the time reversal singlet 
orbits, and a random hopping model gives an accurate description of 
orbit lengths. However, for any finite $K$, the random hopping
approximation fails qualitatively when $N$ is large enough ($N\gg 2^{(2^K)}$). 
\end{abstract}
\begin{keyword}
Gene Regulatory networks; Random boolean networks; Time-reversible Boolean
networks; Cellular automata; 
\end{keyword}
\end{frontmatter}
\newpage

\section{Introduction}
\subsection{Review of dynamical boolean networks}
In recent years considerable effort has been devoted to the study of the
development of complexity in dynamical systems.
Complexity is observed in such different examples as
ecosystems (see \cite{KBR98}),
spin glasses (see \cite{Der87}), and quite
broadly through the
biological sciences\cite{Kau95}.
One thread of activity involves the study of
the behavior of dynamical systems
consisting of $N$ variables $\sigma_t^j$
(the site label $j = 1,\ldots,N$), where each $\sigma_t^j$ is
Boolean, so that it takes on one of two values that we choose
to be $\pm 1$.
The configuration of the system at time $t$ is characterized by
a `state' $\Sigma_t$:
\beq
\Sigma_t=(\sigma_t^1,\sigma_t^2,\ldots, \sigma_t^j, \ldots, \sigma_t^N).
\eeq
The time development is given by saying that
the state at time $t+1$ is a prescribed function
of the state at time $t$, i.e.
\beq
\Sigma_{t+1}= \scr{M}(\Sigma_t) ~.
\label{SKa}
\eeq
The mapping can also be written
\beq
\sigma^j_{t+1} = F^j(\Sigma_t)   \text{\qquad for ~~} j=1, \dots, N
\label{SKb}
\eeq
where the $F^j$ also take on the values $\pm1$.

The number of different states of the system is finite; it is
\begin{equation}
\omega=2^N ~,
\label{omega}
\end{equation}
and the mapping function $\scr{M}$ is independent of time.
Therefore, starting from any state,
eventually the system falls into a cyclic behavior and
follows that cycle forever.

Typically, each of the functions $F^j(\Sigma)$ is picked so that
it depends upon exactly
$K$ distinct input spin variables in the vector $\Sigma$.
A random choice is made to determine which components,
$\sigma^k$ will appear in each $ F^j(\Sigma) $; this fixes the `wiring' of the realization.
Once $K$ and $N$ and the assignment $\sigma^k$'s are fixed, the
each mapping function $ F^j(\Sigma) $
is selected at random from the set
of all Boolean functions of $K$ Boolean variables.
The randomly chosen $N$ sets of $K$ input spin variables and the
functions $F^{j}$ compose a {\em realization}. Given an initial
configuration of the system, a realization will completely define the system's
behavior. The realization is picked at the
beginning of the calculation for the system and remains independent of
time. Since there are $2^{2^K}$ Boolean functions of $K$
Boolean variables and $\left(^{K}_{N}\right)$ different ways to assign
$K$ distinct input spin variables, as $K$ and/or $N$ become large
the number of different realizations is truly huge.

This kind of model is often 
called a {\em Kauffman net} because
Stuart Kauffman\cite{Kau95,Kau84,Kau93} developed a program of study for
generic maps of this type.
Later, this program was extended by Derrida\cite{Der87},
Flyvberg\cite{FK88}, Parisi\cite{BP96,BP97,BP97.2},
and others\cite{Liang,other NK}.
Quantities of interest that have been
studied include the distribution of
cycle-lengths and the number of starting points
which will eventually lead to a given cycle.  (These points form
what is called the  basin of attraction of the cycle.) One calculates the
above quantities by enumeration or by Monte Carlo simulation for each
realization and then averages over realizations with the same values
of $N$ and $K$. 

The Kauffman net is said to be `dissipative'
since several different states may map into one.
Thereby information is lost.

The behavior of Kauffman nets are interesting and surprising.
Large $K$-values produce a
complex time-behavior which closely resembles Parisi's theory of spin
glasses\cite{Der87,PAR,BP97}.
For large
$K$, the cycle lengths grow exponentially with
the number of spins $N$.
Conversely, for $K=0$ or
$K=1$, the cycles tend to be short.
At the `critical' value, $K=2$, typical cycle lengths
grow as a power of $N$, and for large $N$ the
probability of
observing a cycle of length $L$ varies as a power of
$L$. (For a study of critical properties see
Refs~\cite{BP97},~\cite{Liang} and~\cite{Der86b}.)
This three-phase
structure is typical of phase transition problems\cite{Ma76} in which an
ordered and a
disordered phase are separated by a critical phase line.

\subsection{A time-reversible network}
Thus, a great deal is known about the behavior
of Kauffman nets, which can be viewed as a class of generic
dynamical mapping problems.
But not all problems are generic.  For example,
many of the systems
considered in Hamiltonian mechanics are {\em reversible}.
Such systems have the property that
some transformation of the coordinates (for example changing the sign of
all velocities) makes
the system retrace its previous path.
Thus a forward motion and its inverse are equally
possible.
This paper is devoted to a study of the behavior of discrete
reversible maps.

In contrast to a dissipative system, in a finite and reversible dynamical
system every possible state is in exactly one cycle.
Because one and
only one state at time $t$ maps into a predefined state at time $t+1$,
any cycle can be traversed equally well forward or backward.
There are no basins of attraction in this kind of system.
The long-term properties are
then described by giving the number of cycles of length $l$, $N(l)$.

It turns out that for the smaller values of $K$, $N(l)$ is an
oscillating function of $l$ in which the small prime divisors of $l$
play a major role. We shall study this effect in a companion paper. For now, we focus on the gross scaling properties
of $N(l)$, by using a cumulative distribution
\begin{equation}
S(l)=\frac{\sum_{j=l+1}^{\infty}jN(j)}{\sum_{k=1}^{\infty}kN(k)},
\label{survival}
\end{equation}
which gives the probability of finding a cycle of length greater than
$l$ by picking the realizations and the
initial cycle element at random. We call $S(l)$ a ``Survival probability''.

\subsubsection{Definition of the model}
We construct our time-reversible maps
using the method of two time-slices that was
studied for discrete systems
by Fredkin and collaborators.\footnote{The
construction of time-reversible models by using
variables from two different times has a long history.
Imagine doing a calculation
in ordinary classical mechanics using small but discrete
time-steps.
The behavior depends upon the position and velocity of each particle,
but one might wish to do the
analysis in terms of positions alone.
To do this, one works with a state
defined by the positions at two closely neighboring times.
The difference in
position at the two times gives an estimate of the velocity vector.
In this way
one can construct a two slice model of particle behavior.}\cite{TOF90}

As in the usual dissipative Kauffman model,
our basic variable is a list, $\Sigma$, of the
$N$ `spin variables', $\sigma^j$:
\beq
\Sigma=(\sigma^1,\sigma^2,\ldots, \sigma^j, \ldots, \sigma^N).
\eeq

The value of $\Sigma$ is given for each integer value of the time,
and written as $\Sigma_t$.
In our reversible mappings the spin configuration at time
$t+1$, $\Sigma_{t+1}$, depends upon
$ \Sigma_t $  {\em and } $ \Sigma_{t-1} $ according to the rule
\beq
\sigma^j_{t+1}=\sigma^j_{t-1} F^j(\Sigma_t) ~,
\label{model}
\eeq
where the $F$'s are picked exactly as in the dissipative Kauffman net.
Since the
$\sigma$'s take on the values $\pm 1$,
our model can be written in the equivalent form
\beq
\sigma^j_{t+1}\sigma^j_{t-1} = F^j(\Sigma_t) ~,
\label{modelT}
\eeq
which exhibits a quite manifest time reversal invariance.
The models given by
\refe{model} are the subject of this paper.

The information needed to predict future time steps
is called the {\em state} of
the system.
In our case, the state at time $t$, $\scr{S}_t$, is
given by two time slices or {\em substates} $\Sigma_t$ and
$\Sigma_{t-1}$ as
\beq
\scr{S}_t =\binom {\Sigma_{t-1}}{\Sigma_{t} } ~,
\label{state}
\eeq
and the full mapping is of the form
\begin{equation}
\scr{S}_{t+1} =\scr M (\scr{S}_t) ~.
\label{update}
\end{equation}
The history of the system is given by listing the substates in order as 
$$
\Sigma_0, \Sigma_1, \Sigma_2, \cdots, \Sigma_n
$$

If there are $N$ spins, the volume of the state space is
\beq
\Omega=2^{2N} =\omega^2 ~.
\eeq
Since the dynamics are
deterministic, $\Omega$ gives the length of the longest possible cycle.
However, as we shall
see, that length is never attained.

\subsection{Questions to be asked}

\begin{figure}
\centering{\epsfxsize=3.5in
\epsfbox{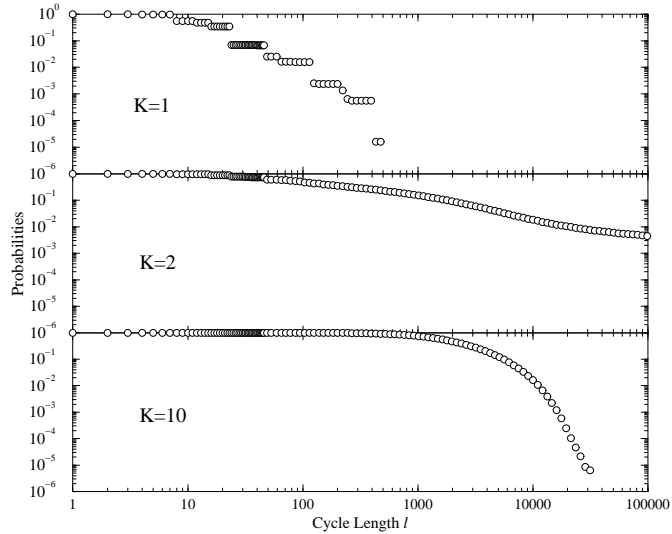}}
\caption{ Plots of probabilities of observing a cycle of length greater
than $l$
as a function of $l$ for
$N=10$  and $K=1,2$, and $N$. These plots are
derived from simulations which average over $10,000$
realizations for each value of $K$. For each realization, one initial
state randomly chosen from the state-space was examined.}
\label{first}
\end{figure}

This paper concerns the distribution of cycle lengths in the
time-reversible models. 
Figure \ref{first} shows plots of probabilities 
of observing a cycle of length larger than $l$
for systems with $N=10$ and various $K$,
averaged over realizations.
For $K=1$ the
cycles are very short; for $K=N$ they have a wide range of
lengths, but the longest ones have length of order $2^N$,
much less than the
number of points in the state space, $2^{2N}$.  For $K=2$,  a wide range of
cycle lengths is seen, including some lengths which
considerably exceed the ones in the $K=N$ system. For smaller values of
$K$, the number of cycles of length $l$ tends to be a strongly
oscillatory function of $l$. We shall discuss this oscillation in a
subsequent publication.

To understand these results for the cycle lengths, we shall
need to understand in some detail the interplay between
the quenched randomness of the system and the time-reversal symmetry.
We will find that the cycles can be divided into two classes;
those that are symmetric under time-reversal and those that
are not.
There are profound differences between the behaviors
of these two types of cycles.

The companion paper will discuss the growth of the `Hamming distance'
between configurations,  defined as the the number of Boolean variables which are unequal in the two
configurations. We shall compare two configurations that
initially differ by a single spin flip, and see how this distance
depends upon the number of
iterations. For small $K$, the growth in the distance may be very slow; for
large $K$ we see much more rapid growth. 

Our simulational data of cycle length distributions
suggest but do not prove the existence of phase transition. However, in
the companion publication we shall see that the behavior of the Hamming
distance can be used to demonstrate convincingly the presence of a type 
of percolation transition at a value of $K$ of about 1.6.  

\subsection {Outline of the rest of paper}
In the next section, we discuss some special features of
time-reversible models and their implications for classifying
different kinds of cycles.
The section after that is devoted to the limiting
cases, $K=0$ and $N$.  Section four describes the structures
seen at intermediate $K$. The appendices cover some more peripheral
issues.
\newpage

\section{Time reversal invariance}

At first sight, it is not obvious that time-reversal invariance should have any
important effect on the distribution of cycle lengths.
However, G. Birkhoff\cite{BIR27} and later John M. Greene\cite{Gre68}
and others\cite{She82} showed one important
mechanism by which the time-reversal process works to set the cycle
length.

\subsection{Symmetry points}

\begin{figure}
\centerline{\epsfxsize=3.5in
\epsfbox{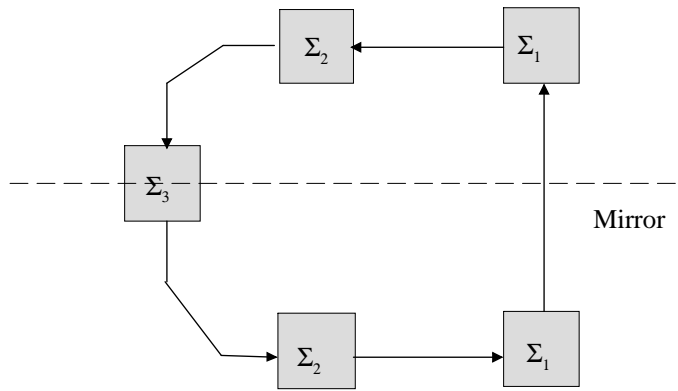}}
\caption{Mirrors produce a closed cycle. The mirrors are at $\Sigma_{3}$
and between the two $\Sigma_{1}$'s. Notice that interchanging the
direction of the arrows gives the same result once again.}
\label{mirrors}
\end{figure}

In our model there are special points in the
phase space which we might call {\em mirrors}.
A mirror produces a
time-reflected motion in the sequence,  for example
\begin{subequations} \label{mirror}
\beq
\dots, \Sigma_3, \Sigma_2, \Sigma_1,
\Sigma_1, \Sigma_2, \Sigma_3, \dots
\label{twin}
\eeq
or, as another example
\beq
 \dots, \Sigma_3, \Sigma_2, \Sigma_1,
\Sigma_0, \Sigma_1, \Sigma_2, \Sigma_3, \dots
\label{sandwich}
\eeq
\end{subequations}
We call the first type (\refe{twin}) a {\em twin} configuration
and the second type (\refe{sandwich}) a
{\em sandwich} configuration. A sketch of the mirrors is given in figure~\ref{mirrors}.
The phase space contains many of these mirrors.
A typical and important cyclic motion is for the sequence of substates
to hit a mirror, be reflected, hit another mirror, be reflected once more,
and thereby be forced into a cyclic behavior.

More explicitly, if we record an orbit of the time-reversible 
model using a sequence of
substates, for example $\dots, \Sigma_{t-1}, \Sigma_t, \Sigma_{t+1},
\dots $, we may call the state $\scr{S}$-value defined as $\binom{\Sigma_{t}}{\Sigma_{t+1}}$ at which
$\Sigma_{t+1}=\Sigma_t$ a twin special point,
and another $\scr{S}$-value a sandwich special point, if $\Sigma_{t+2}$
in the sequence equals $\Sigma_t$. Thus a twin point is any state of the form
\begin{equation}
\scr{S}= \binom{\Sigma} {\Sigma}~,
\label{I}
\end{equation}
so that the cycle will spread out in a palindromic fashion  before and
after this special point in the pattern of
\refe{twin}.
Since the entire volume of the state space is $\Omega=\omega^2$ and
since there are
$\omega$ of these invariant points, the chance that a randomly
chosen point in the state space is a twin point is $1/\omega$.
Correspondingly, a sandwich point appears when there is some
value of
$\Sigma^{(s)}$ for which
\beq
F^j(\Sigma^{(s)})=1
 \text{~~for all~~} j 
\label{sandF}
\eeq	 (see \refe{model}).
A sandwich point gives rise to a sequence of the form of \refe{sandwich}.
The number of sandwich points depends on the realization.
For example,
if $F^j\equiv-1$ for any $j$, then there are no sandwich points at all.
For a given realization, we denote the number of substates
that have the property of \refe{sandF} by $m$, and
we denote each such substate by the symbol $\Sigma^{(s)}_\alpha$;
the $\alpha$ label differentiates between the $m$ different
sandwich substates.
Since for any substate $\Gamma$ a state of the form
$\scr{S}=\binom{\Gamma}{\Sigma^{(s)}_\alpha}$ is a sandwich state,
there must be $m \omega $ of these sandwich states.

Appendix A discusses the properties of the sandwich points. As 
shown there, the average over realizations $\langle m \rangle$ is
unity. When $K=N$ and $N$ is large, $m$ follows a Poisson distribution, 
\begin{equation}
P(m)=\frac{1}{m!e}\qquad\mbox{($N \rightarrow\infty$)}.
\end{equation}
When $K$ is small, there are large realization-to-realization 
fluctuations in $m$,
and the average over realizations of higher powers of
$m$, for example $\langle m^2 \rangle$,
can grow rapidly as a function of $N$. In fact, when $N\gg2^{(2^K)}$,
the vast majority of realizations have $m=0$.

\subsection{Inversions and cycles}
\label{specialpoints}
Two states $\scr{S}$ and $\scr{S}'$ are time reversed images of one
another if $\scr{S}=\binom{\Sigma_1}{\Sigma_2}$
while ${\scr{S}'}=\binom{\Sigma_2}{\Sigma_1}$.
All cycles belong in one of two classes.  The first class, which we call
 {\em special cycles}, each contains at least one pair of time-reversed
images of one another.  The other class, called {\em regular cycles},
contain no such pairs. In Appendix B we show that each special 
cycle of length greater than one contains
exactly two distinct special points. Furthermore, if the points are both
 twins or both sandwiches, the cycle has even length; if they are of different
types, the cycle length is odd.

\subsection {Counting special points
and special cycles}
The arguments in subsection 2.1 imply
that for a given realization
there are $m\omega$ sandwich points and
$\omega$ twin points.
Among them there are $m$ points that are both twin and sandwich
points (consisting of a sandwich substate followed by the same substate),
all of which produce $m$ cycles of length one.
Thus, there are
$(m+1)\omega -m $ distinct special points in the state space.
The total number of special cycles
is the number derived from the points which are
both twins and sandwiches,
$m$, plus the number derived from all the other
special points, $[(m+1)\omega -2 m]/2$, yielding
\beq
\text{number of special cycles} = \frac{(m+1) \omega} {2} ~.
\eeq
The importance of these special points in determining cycle
properties will become clear in the following sections.

\newpage

\section{Limiting cases}
This section discusses the behavior
of reversible Boolean nets for the cases $K=0$ and $K=N$. 

\subsection{$K=0$}
\label{k=0}
When $K=0$ the evolutions of the different $\sigma^j$'s are uncorrelated. Each
$\sigma^j_t$ repeats after either one, two, or four steps. (In contrast, in
the Kauffman net after the first step each of the $\sigma^j$'s remains
constant.) For large $N$, each system is likely to contain at
least one $\sigma^j$ with period four. Hence cycles of period four will
dominate. There will be roughly $N(4)\approx \omega^2/4$ such cycles.

The uncorrelated cycles of the spins produce a hamming distance which
can have value zero or one and has period one, two or four.
 
\subsection{ $K=N$ }
In this part we first review the results for the dissipative Kauffman net
$$
{\sigma^j}_{t+1} = F^j(\Sigma_t) 
$$
and then proceed to discuss
the time-reversible case
$$
{\sigma^j}_{t+1} = F^j(\Sigma_t){\sigma^j}_{t-1} ~.
$$

\label{k=n}
\subsubsection{Dissipative case}
\label{K:net}
The case in which $K$ has its maximum value, $K$=$N$, was first
analyzed for the dissipative Kauffman model by Derrida\cite{Der87}.
This case has the simplifying feature that
a change of a single spin changes the
input of every function $F^j$.
Since the functions are chosen randomly, this means that
every new input configuration $\Sigma_{t}$
leads to an output configuration $\Sigma_{t+1}$ that is picked
at random from the whole phase space,
with its volume
$$
\omega = 2^N~.
$$
This process continues until the time $T$ at which
$\Sigma_{T} = \Sigma_{\tau}$ for some $\tau<T$.
After that, the 
system cycles repeatedly through the sequence
$\Sigma_\tau,\ldots,\Sigma_{T-1}$.

To find the distribution of orbit lengths, we first calculate $p_n$,
the probability that starting
from a randomly chosen initial state at time $t=0$, the
orbit closes at time $n$.
To do this, we define $q_n$ to be
the probability that the cycle remains unclosed after $n$ steps.
The probability that a closure event occurs at time zero is
$p_0=0$, and thus $q_0=1$.
At the time one, the system has a probability $p_1=1/\omega$
of falling into the initial value and a probability
$q_1=1-p_1$ of not doing so.
At time two, there are two possible cycle-closures, since
the new element can be the same as either the zeroth or the first element.
Thus the conditional probability of a closure at time two,
given that the closure event did not occur
at any earlier time, is $2/\omega$.
Similarly, the conditional probability of a closure event
at time $t=n$, given that the system has not closed at time
$t=n-1$, is just $n/\omega$.
Thus the likelihood of a cycle closure at step $n$ is
\begin{subequations} \label{random Kauf}
\beq
p_n= \frac{n}{\omega} q_{n-1}
\eeq
and correspondingly the $q_n$ satisfy
\beq
\label{NKqrecursion}
q_n= \left (1- \frac{n}{\omega} \right ) q_{n-1} ~.
\eeq
\end{subequations}
The solution to \refe{NKqrecursion} with $q_0=1$ is
\beq \label{soln}
q_n= \prod_{j=1}^n (1- j/\omega) ~.
\label{regseq}
\eeq
We shall see that $q_n\ll 1$ unless $n\ll\omega$.
Therefore, we can write
\beq
\ln q_n = \sum_{j=1}^n \ln (1- j/\omega)
\eeq
and expand the right hand side for small $j/\omega$,
yielding\footnote{One can also write
$q_n = \omega^{-n} \omega !/(\omega -n)!$ and expand the
factorials using Sterling's approximation.}
\beq
q_n = \exp[-n(n+1)/2\omega]~.
\eeq
The probability $p_n$
of obtaining a cycle closure at time $t=n$
is then
\beq
p_n =  \frac{n}{\omega}  q_{n-1}
=\frac{n}{\omega}e^{-n(n+1)/2\omega}~ .
\eeq

To obtain $\scr P(L)$, the
probability that a given starting point is in the basin
of attraction of a cycle of length $L$, we note
that a closure event at time $t=n$ yields with equal
probability all cycle lengths up to $n$.
Therefore,
\beq
\scr P(L) = \sum_{n=L}^{\infty}\frac{p_n}{n},
\eeq
which is well-approximated by 
\beq
\scr P(L) & \approx  &  \int_{x=L}^\infty
\frac{1}{\omega} e^{-\frac{x(x+1)}{2\omega}}dx \\
& = & e^{1/8\omega}\frac{\pi}{2}
\left ( {\rm Erf} \left ( \frac{1+2L}{2\sqrt{2\omega}} \right )
\right )~.
\eeq

Note that the probability distribution of cycle lengths is
asymptotically Gaussian (for large lengths), and that 
the length of a typical cycle is of order
$\sqrt{\omega}$\cite{Kau93,BP97}. Derrida
has pointed out\cite{DF86} that the random hopping assumption (or
annealed approximation) is exact for the dissipative Kauffman net with
$K=N$. When $K<N$, the random hopping assumption is no longer exact.
However, the $K=N$ results agree well with the simulational data 
for large but finite $K$.

\subsubsection{$K=N$: Reversible case}
\label{kn_reversible}
As we shall see, the behavior of the reversible model in certain
parameter regimes is also
well-described by a model in which the time-development can be
considered to be random until a closure event occurs.
However, the analysis of $K=N$ limit is more subtle for the
reversible model than for the dissipative case.

{\bf Wrong calculation: leave out special points.}
To illustrate some of 	the complications that arise when we consider
the reversible model, we first
present a naive (and wrong) adaptation to the reversible
system of the argument in subsection \ref{K:net}.
Note that in the reversible models, the state at time t, $\scr{S}_t$,
depends on the spin configurations, or substates, at
{\em two} times, $\Sigma_{t-1}$ and $\Sigma_{t}$.
We once again consider a sequence of states
$\scr{S}_{0}, \scr{S}_{1}, \scr{S}_{2}, \ldots$ and {\em assume} that
the map induces ``random hopping'' through the state space
(each $\scr{S}_j$ chosen with equal
probability from all allowed possibilities).
If the state $\scr{S}_{n}$ happens to be the same as $\scr{S}_{0}$,
then the cycle closes, with the cycle-length
being $L=n$.
Note that at the $n$th step there is only
one possible output that will give closure, $\scr{S}_{n} = \scr{S}_{0}$;
the $n-1$ other values of $\scr{S}_{j}$, $1\le j < n$,
are impossible because each cycle must be traversable
both forward and backward.
Therefore,
if the cycle has not closed in the first $n-1$ steps, the total
number of allowed possibilities for $\scr{S}_{n}$  is $\Omega-n$, of
which only one will give closure.
This argument
yields an estimate for $\rho_n$, the probability of closure at the
$n$th step, given that the orbit has not closed previously:
\begin{equation}
\rho_n= \frac 1  {\Omega-n} ~.
\label{wrongrho}
\end{equation}
Therefore, this estimate implies that $p_l$,
the probability that the orbit closes at the $l$th step, should be
\beq
p_l & =  & \rho_l \prod_{k=1}^{l-1} \left( 1-\rho_k \right)\\
& \approx & \frac{\exp[-l/\Omega]}{\Omega}
\label{wrongp}
\eeq
in the limit of large $\Omega$.

However,   \refe{wrongp} is wrong.
Looking back at figure \ref{first}, one sees that
for $K=N=10$, the
average cycle length is of order $\omega=2^N \approx 10^3$.
However, \refe{wrongp} implies an average
cycle length of order $\Omega=2^{2N} \approx 10^6$.

{\bf A more accurate calculation.}
The problem with \refe{wrongp} is that we have ignored
the role of the special points.
A sequence of substates of the form
\beq
\Sigma_{t_1},\Sigma_{t_1+1},\ldots,\Sigma_{t_1+j},\Sigma^*,\Sigma^*\nonumber
\eeq
is reflected at the twin point and {\em must} continue 
$\Sigma_{t_1+j}$,$\ldots$,$\Sigma_{t_1+1},\Sigma_{t_1}$.
Similarly, a sequence of substates of the form
\beq
\Sigma_{t_1},\Sigma_{t_1+1},...,\Sigma_{t_1+j},\Sigma^*,
\Sigma_{t_1+j+2},\Sigma^* \nonumber
\eeq
is reflected at the sandwich
point. After the first special point has been hit, the orbit
retraces and then continues until a second special point is reached.
Once the second special point is reached, say at time
$t=t^*$, the orbit is reflected again, and
closure in less than $t^*$ additional steps is guaranteed.
Since twin points are hit with probability $1/\omega$ and sandwiches
with probability $m/\omega$ 
at each time step,
this mechanism yields orbit lengths of order $\omega$
rather than the $O(\Omega)$ result of \refe{wrongp}.

In the dissipative case discussed above, when $K=N$, the calculation of
cycle lengths is exact. We have not been able to do that well in the
reversible case.
However, we present here a simple approximation for the
distribution of cycle lengths that
is rather accurate
when $\omega$ is large and $l$ is less than or of the same order as
$\omega$.

We begin by specifying a realization of the network.
Given this realization, we consider a sequence of substates
\beq
G_l =  \Sigma_{0},\Sigma_{1},
\Sigma_{2}, ... , \Sigma_{l}, \Sigma_{l+1}
\label{reg_seq}
\eeq
produced by the map.

We define a {\em regular sequence} to be
one which has neither closed  nor reached a
special point. A regular sequence can be used to construct a part of
either a regular or  a special cycle.
We define the sequence~(\ref{reg_seq}) to be
a sequence of length $l$. Given a regular sequence $G_l$, we may produce a $G_{l+1}$ by evolving $G_l$ for another step. We define the probability $q(l)$ as the fraction of the realizations in which $G_{l+1}$ is also a regular sequence and the probability $\rho(l)$ as the fraction in which it is not.
 
We take $G_l$ as a regular sequence and now
imagine calculating the next substate $\Sigma_{l+2}$. 
As an approximation, assume
that {\em all} $\Sigma_{l+2}$ appear 
with equal probability, $\frac{1}{\omega}$.  
The probability that $\Sigma_{l+2}=\Sigma_{l+1}$
(i.e., $\Sigma_{l+1}$ is a twin point) is $1/\omega$, and
the probability that $\Sigma_{l+2}=\Sigma_l$
(i.e., $\Sigma_{l+1}$ is a sandwich point) is $m/\omega$.
There is also a chance $1/\omega$ that $\Sigma_{l+2}=\Sigma_0$.
In this last case, the orbit will close with no special
points if, in addition, $\Sigma_{l+3}=\Sigma_1$.
Thus, this estimate yields a probability of closure without special
points that is of order $1/\Omega$, as in our naive estimate above.

Therefore, $\rho(l)$,
the probability of a closure event at step $l$,
given that the sequence has not closed previously,
is the sum of two terms:
the probability for closure to a regular sequence
$\rho^R \approx 1/\Omega$, and
the probability of
getting a special point $\rho^S \approx (m+1)/\omega$,
so that
\beq
\rho(l)=\rho^R(l)+\rho^S(l)  \approx (m+1) / \omega~.
\eeq
In addition to ignoring the possibility that $\Sigma_{l+2}$
has already appeared in the sequence,
we have ignored terms of relative order $1/\omega$. This is quite
reasonable for large $N$.

Now we can derive expressions for the number of cycles of
different types.
There are $\Omega$ different starting configurations
for sequences.
We require that the first substate $\Sigma_0$
not be a sandwich substate and that the second
substate $\Sigma_1$ not be equal to $\Sigma_0$,
Therefore, the fraction of sequences of the
form $\Sigma_0,\Sigma_1$ (e.g., $l=0$) that are regular
is $(1-1/\omega)(1-m/\omega) \sim (1-(m+1)/\omega)$.
Each iteration reduces the fraction of
sequences that are regular
by a factor of $1-\rho$, until after $l$ steps we find
that the number of regular sequences,
$N_{RS}(l)$, is (where again we disregard terms
of relative order $l/\omega$ and smaller)
\beq
N_{RS}(l) = \Omega (1-(m+1)/\omega)(1-\rho)^{l}
\approx  \Omega e^{-(m+1)l/ \omega} ~.
\eeq
Since the probability of closure to a regular cycle is $1/\Omega$,
the probability that a randomly chosen point in the phase
space is part of a regular cycle that closes in $l$
steps, $P_R(l)$, is
$P_R(l) = {N_{RS}(l)}/{\Omega }$.
Because each cycle of length $l$ is found by starting
at any of $l$ points on the cycle,
the average number of
regular cycles which close after $l$ steps is
\beq
N_{R}(l) =\frac {N_{RS}(l)}{\Omega l} \approx
l^{-1}e^{-(m+1)l/ \omega}.  \label{NR}
\eeq

A very small
proportion of the points in the state space are,
in fact, parts of regular cycles.
The number which take part in regular cycles of all lengths,
$\text{M}_{R}$, is
\beq
\text{M}_{R} = \sum_{j=0}^\infty j N_{R}(j)
\approx  \frac{\omega}{m+1} ~.
\eeq
This number is indeed much
smaller than the state space volume $\Omega$.

We now turn our attention to the special cycles, which
dominate the state space in this high $K$ limit.
All of them can be found by starting at a special
point and iterating until a second special point
is reached after $l$ steps.
Then the orbit reverses itself and closes after an
additional $l$ steps.
There are three
kinds of special cycles, twin-twin, sandwich-sandwich,
and twin-sandwich.
There are $\omega m$ different ways to choose
the initial point if it is a sandwich point and
$\omega$ ways to choose it if it is a twin point,
so the number of twin-twin cycles and
sandwich-sandwich cycles of length $l$ are:
\begin{subequations}\label{NS}
\beq
N_{tt}(l) = \frac 1 2 e^{-(m+1)l/(2\omega)}   \\
N_{ss}(l) \approx \frac {m^2}{2} e^{-(m+1)l/(2\omega)}~.
\eeq
The factors of two arise in \refe{NS} because each cycle of these
types is found twice by this method.
Similarly, the number of odd (sandwich-twin) special
cycles is
\beq
N_{st}(l) \approx   m e^{-(m+1)l/(2\omega)} ~.
\eeq
\end{subequations}

Note that this estimate for the distribution
of special cycle lengths depends exponentially on $l$, in contrast
to the Gaussian dependence for the dissipative model.

\begin{figure}
\centerline{\epsfxsize=3.5in
\epsfbox{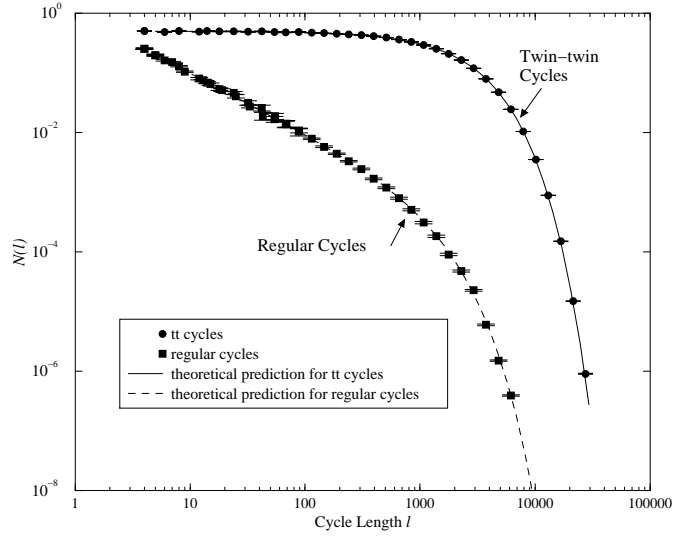}}
\caption{$N(l)$ as a function of $l$, where $N(l)$ is the number of
cycles of length $l$, averaged over realizations.
The curves are the results of the random hopping approximation 
(equations (\ref{NR}) and (\ref{NS}))  and the points
are simulational data. Simulation results are plotted after being averaged
over 10,000 realizations with $K=N=10$ and $m=0$. }
\label{test1K=N}
\end{figure}

To check these conclusions we plot in figure
\ref{test1K=N}  $N(l)$, the number of cycles of length $l$ as a
function of $l$ averaged over realizations. The realizations used all
had $m=0$. The theoretical estimates (equations~\ref{NR} and~\ref{NS}) agree
very well with the numerical results. 

{\bf Average over $\mathbf{m}$.} 
Our results of equations (\ref{NR}) and (\ref{NS}) depend upon the
number of sandwich special points, $m$.
In this section, we shall denote averages over $m$ by
$< \cdot >$.
In Appendix A we
show that for $K=N$, the probability distribution for $m$
is a Poisson distribution
\beq
\rho
(m)=\frac{e^{-\lambda}\lambda^{m}}{m!} \text{~~with~~}\lambda=1 ~.
\label{Poisson}
\eeq
Averaging \refe{NR} using the weight defined
in \refe{Poisson} gives that the realization average of the
number of regular cycles of length $l$, $<N_{R}(l)>$, is
\begin{subequations} \label{NSA}
\beq
<N_{R}(l)> =
l^{-1}\exp\left[(e^{-l/\omega}-1)
-l/\omega\right] ~.
\eeq
Similarly, for the various kinds of special cycles
\beq
<N_{tt}(l)> & = &  \frac 1 2 \exp\left[(e^{-l/(2\omega)}-1)
-l/(2\omega)\right] \\
<N_{ts}(l)> &=& \exp\left[(e^{-l/(2\omega)}-1)
-l/\omega\right]  \\
<N_{ss}(l)> &=&  \frac{e^{-l/(2\omega)}+1}{2}
\exp\left[(e^{-l/(2\omega)}-1)  -l/\omega\right] ~.
\nonumber  \\
\eeq
\end{subequations}

\begin{figure}
\centerline{\epsfxsize=3.5in
\epsfbox{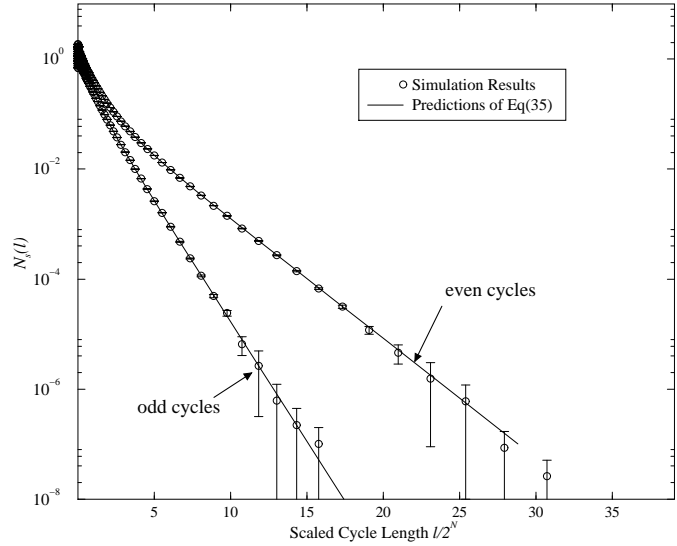}}
\caption{The average number of special cycles of length $l$, $N_S(l)$,
plotted against scaled cycle length $l/2^N$. The result is averaged 
over 10,000 randomly selected
realizations for $K=N=10$ and hence is an average over $m$. For each
realization all special cycles were enumerated. The symbols are
the results of the simulations. The theoretical results for $N_S(l)$
obtained from \refe{NSA} are included as solid lines. }
\label{test2K=N}
\end{figure}

To test \refe{NSA} against simulations, in figure \ref{test2K=N} we
plot $N_S(l)$, the number of special cycles averaged over 
realizations, against $l$. 
As discussed in
subsection~(\ref{specialpoints}), special cycles formed by two twin points
or two sandwich points have an even length, while the ones with one
twin point and one sandwich point have an odd length. Therefore,
$N_S(l)=<N_{ss}(l)>+<N_{tt}(l)>$ if $l$ is even and $N_S(l)=<N_{ts}(l)>$ if
$l$ is odd. We find that $N_S(l)$ oscillates because of this difference
between even cycle lengths and odd cycle lengths, leading to a
two-branch structure in $N_S(l)$. This structure was indeed observed in
our simulations. 

The random hopping analysis of the $K=N$ situation gives a
Hamming-distance behavior which is exactly the same in the dissipative~
\cite{Kau93} and reversible systems. In both cases, the systems start
from a pair of configurations which differ in the value of one spin at
time $t$. By the next time-step, the configuration will be random, so
that half the spins will be ``wrong''. Thus the Hamming distance
immediately comes to the value $N/2$, and stays there.
\newpage

\section{Intermediate $K$}

\label{OtherKs}

Section 3 outlines both qualitative and quantitive pictures of the
limiting cases $K=0$ and $N$. In this section, 
we will discuss the system's 
behavior for intermediate $K$ values. As $K$ is increased, 
there is an evolution
from dynamically independent clumps of spins to a situation in which there is
random hopping over the whole state-space.
 
\begin{figure}
\centerline{\epsfxsize=4.5in
\epsfbox{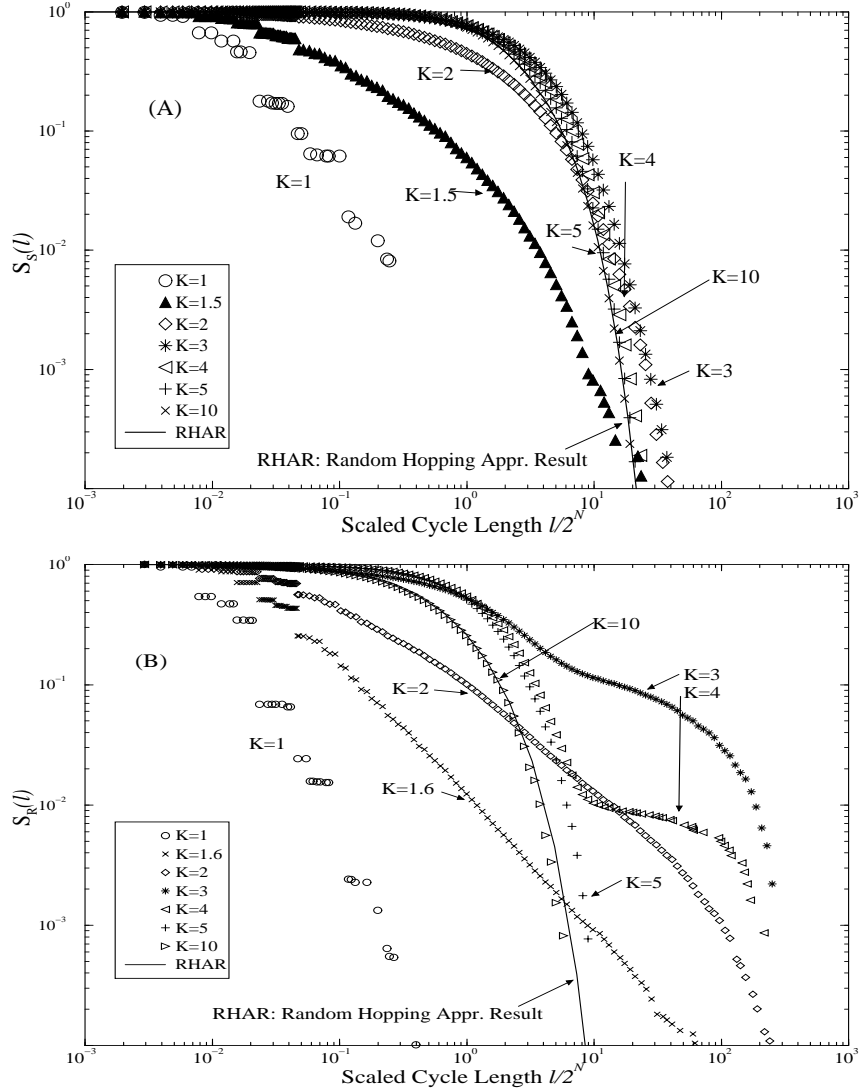}}
\caption{The plot of survival functions of scaled cycle length
$\hat{l}=l/2^N$ for both special and regular cycles for various $K$
values and $N=10$. (A) shows the survival function for special cycles,
$S_S(l)$, and (B) shows the survival function for regular cycles,
$S_R(l)$. The functions were averaged over 25600 realizations
for each $K$ value. The survival functions converge to the random
hopping case for larger $K$ values. However a qualitative theory  
is still needed to predict the function forms for small $K$. }
\label{fin141grf}
\end{figure}

Figure~\ref{fin141grf} shows numerical
results of $S_S(l)$ and $S_R(l)$, the survival functions for special
cycles and regular cycles separately, for various intermediate $K$
values. Recall that a survival function of cycle length $l$ is 
defined to be the probability
of observing cycles with length greater than $l$ 
(see~\refe{survival}). Noninteger values of $K$ are produced by
mixtures of spins with the neighboring integer $K$-values.   
The survival functions
for special cycles, $S_S(l)$, as shown in part (A), shows no
surprising structure. As $K$ decreases from 2, the observed cycle
lengths get shorter and shorter. For $K\geq 3$, the survival
probabilities approach in a
uniform manner the result of~\refe{NSA}.
The regular cycles are different. For $K>5$ we see a uniform approach to
the $K=N$ result. At $K=1.6$, there seems to be a power law behavior
with $S_R(l)\sim l^{-1}$. Then for larger values of $K$, $S_R(l)$ seems to
decrease faster than power law. Finally, $K=3$ and $K=4$ show a
remarkable bump at large values of $l/2^N$, indicating that there is a
new process going on at large $l$. 

\begin{figure}
\centerline{\epsfxsize=3.5in
\epsfbox{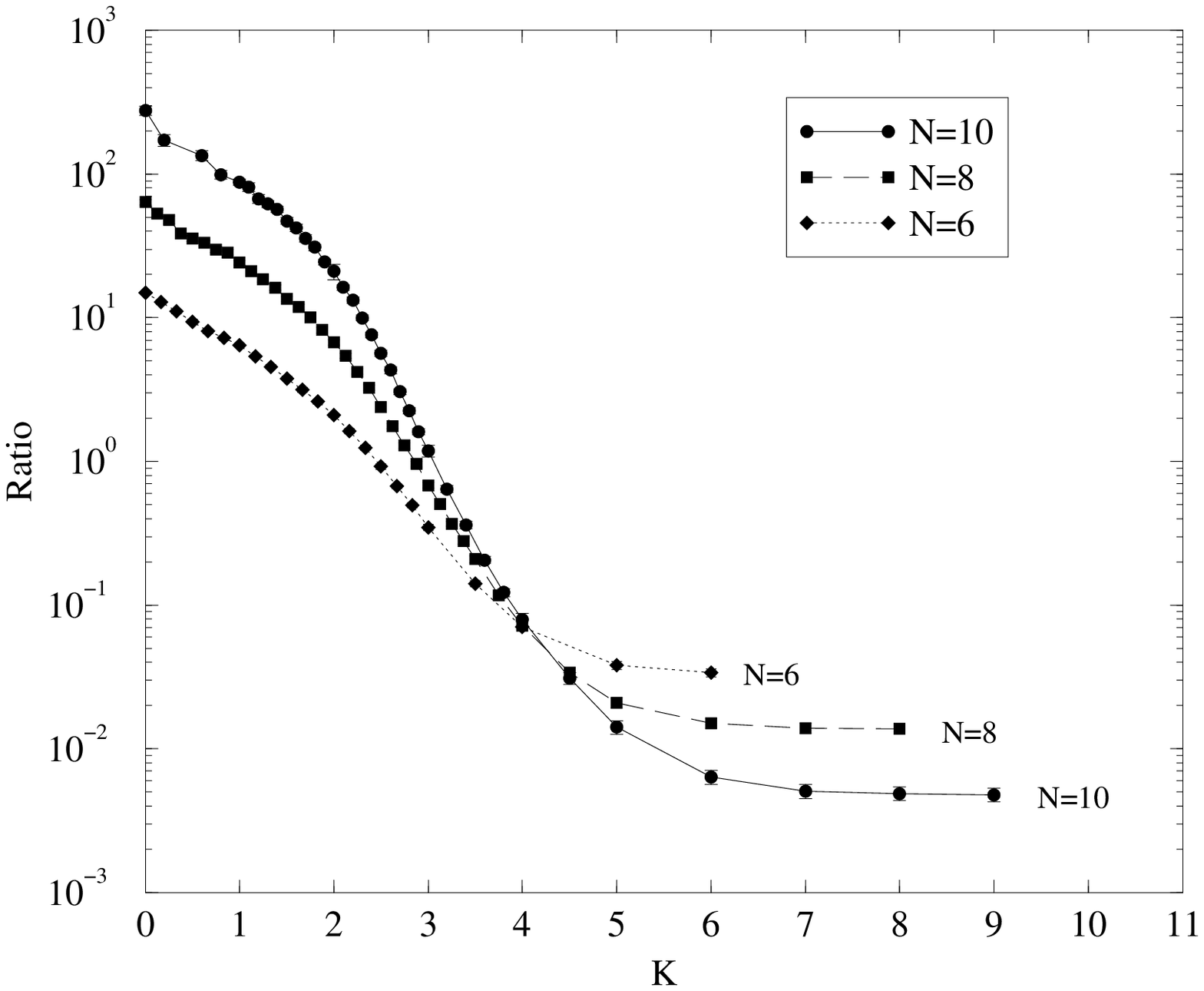}}
\caption{The ratio of the number of regular cycles to the number of
special cycles for N=6, 8, and 10. Both the numerator and denominator 
are averaged over 25600 realizations. At small $K$ there 
are many more regular cycles than special cycles; at large $K$ most 
of the cycles are special.}
\label{ratio}
\end{figure}

The difference between part (A) and (B) implies that the relative
importance of the two pieces of cycle-closing mechanism depends strongly
on K. Moreover, the relative numbers of
regular and special cycles are also strongly $K$-dependent. This point is
illustrated in figure~\ref{ratio}, which shows the ratio of the number
of regular cycles to that of special cycles as a function of $K$.
For small $K$, several subsets of the spins may become uncoupled, 
and most cycles are regular cycles. Conversely special cycles
are more likely for large $K$.

\subsection{Anomalously Long Regular Cycles for Intermediate $K$ Values}

\begin{figure}
\centerline{\epsfxsize=3.5in
\epsfbox{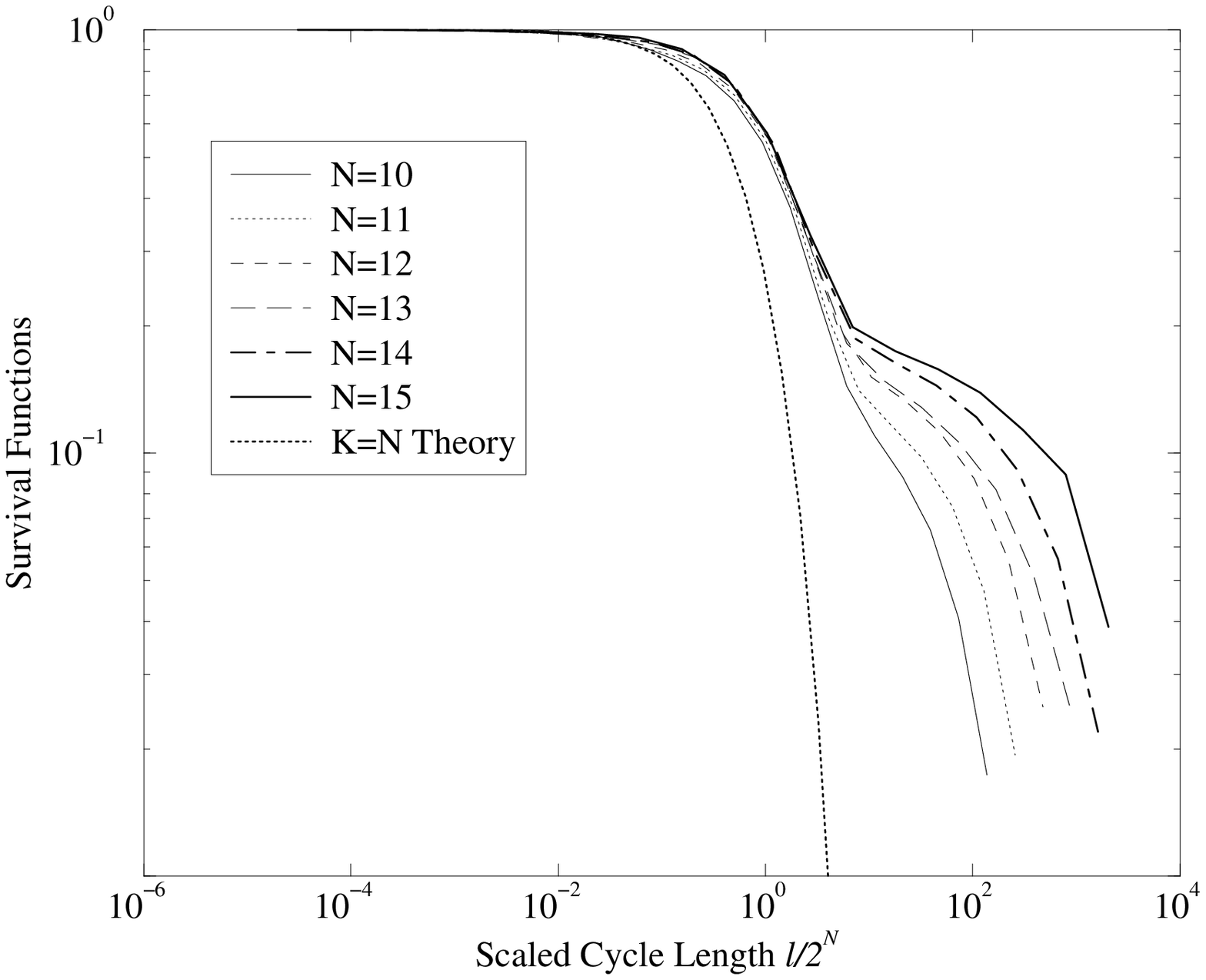}}
\caption{Plot of $S_R(l)$ against $\hat{l}\equiv l/2^N$ for
$K=3$ and various $N$. The drop of $S_R(l)$
corresponding to the relatively short cycles scale as $2^{N}$, 
which supports the argument that they are due to special points.}
\label{fin81grf}
\end{figure}

Notice the bump in figure~\ref{fin141grf} part (B) 
for $K=3$ and $K=4$ in the region
$1\ll\hat{l}\ll\omega$. This bump arises from a group of regular cycles
which are anomalously long. More careful study shows that in fact regular
cycles split into two groups, which scale differently. To demonstrate this
effect, we plot in 
figure~\ref{fin81grf} the survival functions $S_R(l)$ for $K=3$ and for a
range of $N$ values against the cycle length $l$ normalized by $2^N$.
Manifestly the distribution of short cycles
scales with $\omega=2^{N}$. This scaling can be
explained in the following way. There are $(m+1)2^N$ special points in
the state-space of the system, as discussed in section 2.3.
To obtain a regular cycle, the system must not
hit any of the special points. The probability of hitting a special
point $\sim 1/2^N$, so the lengths of the
regular cycles scale as $2^{N}$. For a more explicit and quantitative 
reasoning, we may appeal to the regular cycle result based on the
random hopping assumption described by \refe{NSA}, which provides a
reasonably accurate approximation of the system
when the system is sufficiently well-connected. Note that in \refe{NSA}, the
regular cycle distribution clearly scales as $2^{N}$. 

While we have presented an intuitive explanation 
for the scaling of the relatively
short regular cycles, one may still wonder why there is
a population of anomalously long cycles and why it appears and disappears as
$K$ increases. To attack these issues, a number of realizations which
produce extra long cycles were studied and similarities among them were
observed. In general, a realization and the initial configuration 
have to satisfy the following two conditions to be able
to generate an anomalously long cycle: The realization should not have
any sandwich points, thereby annulling the mechanism to close a cycle
using them, and the functions assigned to the spins 
together with the choice of 
initial configurations should prevent a system from hitting a twin
point. For small $K$ values, one can easily find realizations without
any sandwich points. For instance, if any spin is assigned the function
that is ``-1'' for all inputs, then the
realization has no sandwich points. In fact, in Appendix A, we
prove that almost all realizations have no sandwich points for
finite $K$ and large enough $N$ ($\gg 2^{(2^K)}$). 

The twin points in the cycle could 
be avoided in many ways. For example, when a
function assigned to a spin equals to the constant $+1$, if this spin
starts from $\sigma_{0}=+1$ and $\sigma_{1}=-1$, it will follow the progression
of $+1$, $-1$, $+1$, $-1$, ... , and the system never hits a twin
point since $\sigma_{t}\neq\sigma_{t+1}$ always holds. Since there is a
probability $1/2^{(2^K)}$ that any given spin is assigned the function
``$+1$'', and a probability $1/2$ that this spin starts with
$\sigma_0=-\sigma_1$, the probability $p_2$ that at least one spin is in the
``$+-+-$'' progression is 
\begin{eqnarray}
p_{2}&=&1-(1-(\frac{1}{2})(\frac{1}{2^{(2^K)}}))^N\nonumber\\
&\approx&1-\exp(-\frac{N}{2^{(2^K-1)}})\qquad (2^{(2^K)}\gg 1).
\end{eqnarray}
Thus, when $N\gg 2^{(2^K)}$, almost all starting
configurations and realizations will not hit a twin point.

Another way to avoid twin points consists of two spins that are assigned 
identical inputs and functions. In this
case, it is possible to choose an initial configuration for these two
spins, which will stop the system from forming a twin
point. For instance, a system in which spin 1 and spin 2 are assigned
identical input spins and functions starting from
$\sigma^{1}_{t=0}=\sigma^{1}_{t=1}=1$ and $\sigma^{2}_{t=0}=-1$,
$\sigma^{2}_{t=1}=1$ can never hit a twin point. It is also easy to
prescribe certain simple progressions for a few spins and stop 
the system from hitting a twin point. For
example, two spins both with period $3$ evolving following the pattern
\begin{eqnarray}
\sigma_{0}:++-++-++-\cdots\nonumber\\
\sigma_{1}:-++-++-++\cdots\nonumber
\end{eqnarray} 
can prevent the system from forming a special cycle. 

The mechanisms preventing the system from hitting a
twin point always appear to be related to some small piece of the system
that evolves independently from other parts of the system. The couplings are
such that this piece is not affected by anything outside
itself (though in general it can and does affect the rest of the
system). 
We call a piece like this a local structure. 
Local structures are to be discussed in detail in the
next paper in this series. Note that local structures are very
unlikely for sufficiently large $K$ values, where the spins of the
system are quite correlated, unless $N$ is enormous. When $K$ is small, 
local structures occur much more frequently. When many a local 
structures are present, almost all the realizations and
initial configurations have no special points. Thus the presence of
local structures leads to very long regular cycles. Since for a given
$N$ the local structures become less probable as $K$ increases, it is natural 
to find the number of regular cycles decreases as $K$. 

\begin{figure}
\centerline{\epsfxsize=3.5in
\epsfbox{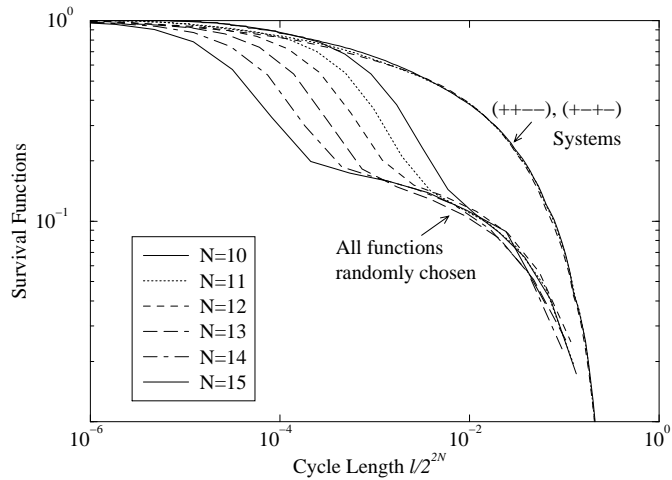}}
\caption{Survival functions of regular cycle lengths in systems with all
functions chosen at random as well as systems with one spin assigned the
function $-1$ and a second spin assigned the function $+1$ and initial
condition $(+1, -1)$, plotted against $l/2^{2N}$ 
for various $N$ and for $K=3$. As expected,  the lengths of anomalously
long cycles in the systems with randomly functions, as well as 
all the cycles in the $(+,+,-,-), (+, -, +, -)$ 
systems, scale approximately as $2^{2N}$.}
\label{fin91grf}
\end{figure}

Section \ref{kn_reversible} presents a naive 
theory of hopping in state space that
ignores the role of the special points and predicts that the distribution
of orbit lengths in a system of $N$ spins should scale as $2^{2N}$. 
This naive theory fails qualitatively when $N=K$ because the special
points induce orbit closure in order $2^N$ steps.  However, if local
structures prevent the system from ever reaching a special point, then the
mechanism for closing the orbits in order $2^N$ steps does not operate and
it is plausible that the typical orbit lengths will be much longer than
$2^N$.  We test this idea by calculating orbit lengths in systems with one
spin assigned the function $-1$ (so it follows the sequence $(+1, +1, -1,
-1, +1, +1, \cdots)$) and a second individual spin assigned the
function $+1$ and the initial condition $(+1,-1)$ (so its evolution is
$(+1, -1, +1, -1,\cdots)$). Such $(+,+,-,-), (+, -, +, -)$ systems can never
reach a special point.  As figure~\ref{fin91grf} demonstrates,
 in these systems orbit lengths grow with $N$ much faster than $2^N$;
the numerical results are consistent with $2^{2N}$ scaling. 

We believe that the orbit lengths in systems with $N \gg 2^{2^K}$ and
randomly chosen functions cannot grow faster than
$2^{2N(1-\epsilon(K))}$, where $\epsilon(K)$ is of order $2^{-2^K}$. This
is because of order $N2^{-2^K}$ spins have input functions $1$ or $-1$ and
hence cycle with periods $1,$ $2$, and $4$.  More generally, we expect
interesting crossover phenomena to occur when $K$ is both large and of the
order of $log_2 (log_2 N)$.  Even for $K=3$, simulating systems with large
enough $N$ to permit numerical exploration of these effects is
computationally prohibitive.

\subsection{Average cycle length versus $N$ for different values of $K$}

\begin{figure}
\centerline{\epsfxsize=3.5in
\epsfbox{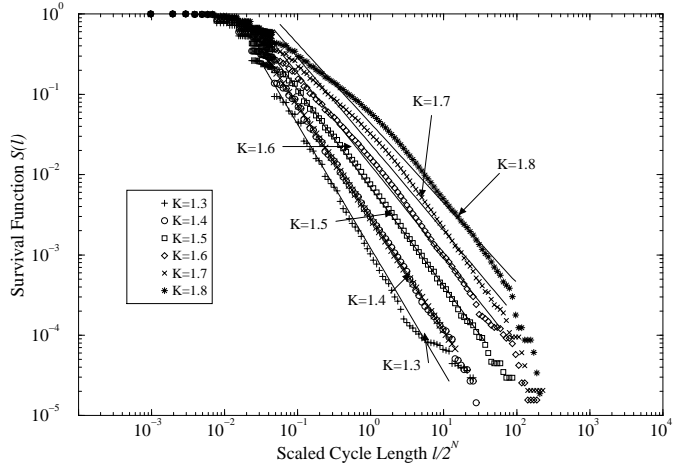}}
\caption{Survival functions of regular cycle lengths 
plotted against $l/2^N$ for various $K$ values and for $N=10$. The
solid lines are the power law fits for the
survival functions. By inspection one can see for $K\in[1.4, 1.7]$ the
survival functions follow power law fairly well for three decades. }
\label{k1-2}
\end{figure}

When $K$ is large, the random hopping approximation works well and the
cycle length distribution scales as $\omega=2^N$. 
Figure~\ref{fin141grf} demonstrates that the distribution of cycle
lengths does not change dramatically while K decreases until $K$ 
is quite small. Therefore, it is reasonable to expect the average cycle length
to increase exponentially with $N$ when $K$ is large. On the other hand,
when $K=0$, the average cycle length is bounded above by 4. For
$K=1$, our simulations and analytic arguments indicate that it scales as $\log{N}$; the
results will be presented in the companion publication. For values of
$K$ in the range [1.4, 1.7], the survival functions $S(l)$ decay roughly
as a power law over three decades in cycle length $l$, as shown in 
figure~\ref{k1-2}. This result is consistent with the presence of a
phase transition. The companion paper on the behavior of the Hamming
distance presents more compelling evidence for a phase transition at
$K\approx 1.6$.

\newpage

\section{Summary}
This paper addresses the dynamics of a Boolean network model of $N$
elements with $K$ inputs with
time-reversible dynamics. We present the general setup of the model and
introduce the concept of special points and the distinction between special
cycles and regular cycles. The relation between special points and
the properties of the cycles is demonstrated. We show that the numbers 
of special points and special
cycles for each realization are proportional to $\omega\equiv 2^N$,
where $N$ is the number of variables in the system. 

We determine the probability distribution of cycle lengths as well as
the survival functions. In limiting case $K=0$, the cycle length is
bounded above by 4 and the probability that a cycle length is 4 
approaches 1 as $N$ increases. For $K=N$, within a random hopping 
approximation we calculate the survival functions for regular cycles and
for special cycles, which agree with simulational data extremely well. 

Finally, we present the simulational results for survival functions for
intermediate $K$ values. A population
of anomalously long regular cycles scaling as $2^{2N}$ is found for
small $K$ values and
explained based on the notions of special points and local structures. 
The correlation between typical cycle length and the K values
of the networks is studied, and we find that the typical cycle length increases
logarithmically with $N$ when $K<1.4$, exponentially when $K>1.7$, and
following a power law when $K$ falls in between; these results are
compatible with  the presence of a phase transition for $K$ somewhere in
the range of  $[1.4, 1.7]$.
\newpage

\appendix

\section*{Appendix A: Some statistical properties of sandwich points}
\label{ms}

The text discusses twin and sandwich symmetry points
which induce closure of orbits in the reversible Kauffman model.
Each type arises when the substate
$\Sigma_t$ is such that each of the functions
of these inputs takes on a target value.
For a twin point, $\sigma_{t+1}^j=\sigma_t^j$,
so the target function is
different at every time step.
For a sandwich point, for all $t$ the target function
is $F^i=1$ for every $i$.

To calculate orbit lengths, we need to compute the probability that
a symmetry point of either type occurs at each time $t$.
For a given realization of couplings, the number of
substates for which the functions take on
a particular value can vary.
Because the target value for the twin point is different for
different substates,
whereas the target values for the sandwich points are the same
at all times,
the statistics of the two types of symmetry points
are different.

The process we consider is one in which couplings
and an initial condition are chosen, and then the system
evolves in time.
We assume that this time evolution yields a random sampling
of all possible spin configurations or substates.
At each time $t$ we examine whether a symmetry point of
either type has been reached.
Let $m$ be the number of substates for which $F^j=1$
for all $j$ (the criterion for a sandwich
point), and $m_t$ be the number of spin configurations
for which each function takes on the target value for a
twin point at time $t$.
At a time $t$, the probability of being at a twin
point is $m_t/\omega$, whereas the probability of being at
a sandwich point is $m/\omega$.
On average it takes
many trials before a special point is reached; the probability
of having observed a sandwich point after $T$ trials
$\approx \sum_{t=1}^T m_t/\omega \approx \langle m \rangle T/\omega$
= $T/\omega$, where $\langle \rangle$ is the 
average over realizations,
whereas the probability of having observed a sandwich symmetry point
is $mT/\omega$.

We wish to calculate the probability that a randomly chosen
realization has a given value of $m$.
Now each output takes
on a given value with probability $1/2$, so on
average the probability
that $N$ outputs all have given target values is
$(1/2)^N$, implying that the realization average
$\langle m \rangle = 2^{-N}$.
However, if one of the functions happens to be $F^i = -1$,
then clearly there are no sandwich points.
We wish to find $P_K(m)$,
the fraction of all possible realizations
of the couplings for a given $K$ that yield each value of $m$.

First we consider $K=N$.
This case is particularly relevant because when $K$ is large,
essentially all the orbits close because of the symmetry points.
Here, the functions can be viewed as mapping a given
input substate into a randomly chosen output substate.
Since there are $\omega =2^N$ possible output substates, of which
one is the target, each input configuration has
a probability $1/\omega$ of having its output be the target.
$P_{K =N}(0)$,
the probability that no input configurations has as its
output the target configuration, is
$(1-2^{-N})^{2^N}$, which, as $N \rightarrow \infty$,
approaches $1/e$.
The probability that exactly one of the $2^N$ different inputs
yields the target output is
$(2^{N}) (2^{-N})(1-2^{-N})^{2^N-1} \rightarrow 1/e$.
Similarly, $P_{K =N}(m)$,
the probability that exactly $m$ of the
$2^N$ different inputs yields the target output is
\begin{eqnarray}
P_{K =N}(m) & = &
\frac{\left (2^N\right )!}{m!(2^N-m)!}\left(2^{-N}\right )^m
\left(1-2^{-N}\right )^{2^N-m}
\nonumber \\*
& \rightarrow & \frac{1}{m!e}
\qquad\mbox{($N \rightarrow \infty$, $m\ll 2^N$)}.
\end{eqnarray}
Thus, $P_{K =N}(m)$ is a Poisson distribution.
Since $P_{K =N}(m)$ falls off quickly
as $m$ gets large, clearly it is consistent to assume
that $m \ll 2^N$.

However, when $K$ is finite and $N$ is large enough,
we expect the behavior of $P_K(m)$ to differ qualitatively from
the $K=N$ result.
We expect that almost all configurations will have
$m=0$ for any finite $K$ as $N \rightarrow \infty$.
We have argued before that if in a realization a spin is assigned a
function that equals to constant -1, the realization has no sandwich
point. Also, if two spins are assigned functions $F^1$ and $F^2$ such
that 
$$
F^1+F^2=0
$$  
for all inputs, there can be no sandwich point for the
realization. There are many other possible mechanisms that lead to
$m=0$. Clearly, the probability that a realization has at
least one spin function of -1 bounds below the
probability that it has no sandwich point. Among $2^{(2^K)}$
possible functions that can be assigned to one spin, one is $-1$ for all
inputs. Assuming that all functions are equally likely to be picked  
and that the function choices for different spins are independent, 
the probability that no spin is assigned the constant function -1 is
$$
(1-\frac{1}{2^{(2^K)}})^N\approx\exp(-\frac{N}{2^{(2^K)}})~~(N\gg 2^{(2^K)}).
$$
Thus, $P_K(m=0)$ is bounded by the probability that the realization has
at least one function that is $-1$; or $P_K(m=0)\geq (1-\exp(-\frac{N}{2^{(2^K)}}))$. This result implies that whenever $K$ is finite,
in a large enough system sandwich points cause
orbit closure only in a vanishingly small fraction of
realizations.
However, when $K$ is not small,
realizations with sandwich points are rare only when
$N$ is enormous (when $N \gg 2^{2^K}$).

\section*{Appendix B: Relation between special points and cycles}

Here we prove the results used in section 2 that 1) each special cycle
 contains exactly two special
points, 2) that cycles with two special points of the same kind have
even cycle lengths, and 3) cycles with different kinds of special points
have odd lengths.

To prove that each special cycle contains two and only two special
points, we first consider a cycle of even length $2n$,
$$
\Sigma_0, \Sigma_1, \ldots, \Sigma_{2n-1}.
$$
Suppose there is a twin point in the cycle. By relabeling the cycle, 
we can get 
$$
\Sigma_{n-1}=\Sigma_{n}
$$
by definition. Now $\Sigma_{n-1-t}=\Sigma_{n+t}$ since the cycle
is time reversible, so that $\Sigma_0=\Sigma_{2n-1}$ when we take
$t=n-1$. Thus we find another twin point in this even-length cycle. If
there is a sandwich point at $n$, then 
$$
\Sigma_{n-1}=\Sigma_{n+1}
$$
and $\Sigma_{n-1-t}=\Sigma_{n+1+t}$, thus
$$
\Sigma_{1}=\Sigma_{2n-1}.
$$
Here we find another sandwich point at 0. Similarly, when the cycle is
odd, we will find a twin point in the presence of a sandwich point, and
vice versa.

By now we have proven that if there is a special point in the cycle,
then there has to be another. The statement that the cycle length
being odd or even depends on whether the special points are of the same
kind, is also clear from the above argument.

To finish the proof, we need to demonstrate 
that no orbit can contain more than
two special points.
Assume that an orbit of length $L$ with more than two special
points exists.  Choose the origin of
time so that
$\Sigma_{L-j}=\Sigma_j$ (one does this by
placing a sandwich substate at $t=0$ or placing twin
substates at $t=1$ and $t=L$), and let $P$ be
the smallest value for which
$\Sigma_{P-n}=\Sigma_{P+n}$ for all $n$\footnote{If the
special point at $P$ is a twin point, then $2P$ is odd, otherwise it is even.};
by assumption, $P < L/2$.
Then we must have simultaneously $\Sigma_{L-j}=\Sigma_j$ and
$\Sigma_{P-n}=\Sigma_{P+n}$.
Letting $j=P-n$ yields
$\Sigma_{L-P+n} = \Sigma_{P-n} = \Sigma_{P+n}$.
Letting $q=P+n$, we obtain
$\Sigma_{L-2P+q}=\Sigma_{q}$.
Thus the orbit period is $L-2P$, which is strictly less than
$L$, so we have a contradiction.

We can also show that these two special points 
must be different from one another.
Suppose not.  Consider an orbit of length $L$, and
choose the origin of time so that $\scr{S}_0$ and $\scr{S}_P$ are the
same special point, with, by assumption, $P<L-1$.
Applying the map yields $\scr{S}_{j } = \scr{S}_{(P+j) }$
for any $j$, so the orbit repeats after $P$ steps.
But this contradicts the assumption that $L$ is strictly
greater than $P$.
\newpage

\end{document}